\documentclass[bimj,fleqn]{w-art}
\usepackage{times}
\usepackage{w-thm}
\usepackage[authoryear]{natbib}
\theoremstyle{plain}

\theoremstyle{definition}

\usepackage[]{graphicx}
\chardef\bslash=`\\ 

\usepackage{amsfonts, amsmath, mathrsfs, bm}

\newcommand{\convergeto}{\overset{d}{\longrightarrow}}
\def\T{{ \mathrm{\scriptscriptstyle \top} }}
\newcommand{\bas}{\begin{eqnarray*}}
\newcommand{\eas}{\end{eqnarray*}}
\newcommand{\ba}{\begin{eqnarray}}
\newcommand{\ea}{\end{eqnarray}}
\newcommand{\bit}{\begin{itemize}}
\newcommand{\eit}{\end{itemize}}
\newcommand{\ben}{\begin{enumerate}}
\newcommand{\een}{\end{enumerate}}
\newcommand{\e}{ { \mathbb{E}}}
\newcommand{\var}{ { \mathbb{V} {\rm ar}}}

\newcommand{\pr}{{\rm pr}}

\newcommand{\bA}{ {\bm A }}
\newcommand{\bbeta}{ {\bm \beta} }

\newcommand{\btheta}{ {\bm \theta}}
\newcommand{\bzero}{ {\bf 0}}

\newcommand{\bO}{ {\bm O}}

\newcommand{\bX}{ {\bm X}}

\newcommand{\bx}{ {\bm x}}

\newcommand{\rV}{ { \rm V}}

\newcommand{\bW}{ {\bf W}}

\newcommand{\bV}{ {\bf V}}

\usepackage{color}

\usepackage{algorithm}
\usepackage{algorithmic}

\usepackage{textcomp}
\usepackage{latexsym}
\usepackage{amsmath}
\usepackage{amssymb}
\usepackage{arydshln}

\usepackage{pmat}
\usepackage{amsfonts, amsmath, mathrsfs, bm}
\usepackage{float, multirow, booktabs, threeparttable}
\usepackage{array}
\usepackage{textcomp}
\usepackage{verbatim}

\newcommand{\PreserveBackslash}[1]{\let\temp=\\#1\let\\=\temp}
\newcolumntype{C}[1]{>{\PreserveBackslash\centering}p{#1}}
\newcolumntype{R}[1]{>{\PreserveBackslash\raggedleft}p{#1}}
\newcolumntype{L}[1]{>{\PreserveBackslash\raggedright}p{#1}}

\usepackage{lineno,hyperref}
\modulolinenumbers[5]

\begin{document}

\Volume{}
\Issue{}
\Year{2021}
\pagespan{1}{}

\title[EL estimation for abundance from one-inflated capture-recapture data]{Semiparametric empirical likelihood inference for abundance
\\from one-inflated capture--recapture data}

\author[Liu {\it{et al.}}]{Yang Liu\inst{1}}
\address[\inst{1}]{KLATASDS - MOE,
School of Statistics,
East China Normal University,
Shanghai 200062, China}

\author[]{Pengfei Li\inst{2}}

\address[\inst{2}]{Department of Statistics and Actuarial Science,
University of Waterloo,
Ontario N2L 3G1, Canada}

\author[]{Yukun Liu
\footnote{Corresponding author: {\sf{e-mail: ykliu@sfs.ecnu.edu.cn}}}
\inst{, 1}}
\author[]{Riquan Zhang\inst{1}}

%

\Receiveddate{zzz} \Reviseddate{zzz} \Accepteddate{zzz}

\begin{abstract}
Abundance estimation from capture--recapture data
is of great importance in many disciplines.
Analysis of capture--recapture data is often complicated by
the existence of  one-inflation and heterogeneity problems.
Simultaneously taking these issues into account,
existing abundance estimation methods  are usually  constructed
on  the basis of conditional likelihood (CL)  under one-inflated zero-truncated count models.
However, the resulting  Horvitz--Thompson-type estimators  
may be unstable, and the resulting Wald-type confidence intervals
may exhibit severe undercoverage.
In this paper, we propose a semiparametric empirical likelihood (EL) approach to abundance estimation
under one-inflated binomial and Poisson regression models.
We  show that the maximum EL estimator for the abundance
follows an asymptotically normal distribution
and that the EL ratio statistic of abundance
follows a limiting chi-square distribution with one degree of freedom.
To facilitate  computation of the EL method,
we develop an expectation-maximization (EM) algorithm,
and establish its appealing convergence property.
We also propose a new score test for the existence of one-inflation
and prove its asymptotic normality.
Our simulation studies indicate that compared with CL-based methods,
the maximum EL estimator has a smaller mean square error,
the EL ratio confidence interval has a remarkable gain in coverage probability,
and the proposed score test is more powerful.
The advantages of the proposed approaches are further demonstrated
by  analyses of   prinia data from Hong Kong and
drug user data from Bangkok.
\end{abstract}

\maketitle                   







\section{Introduction}
\label{s:intro}

Knowledge of the abundance of species or the sizes of hidden and elusive populations is
of importance in ecology, epidemiology, public health, and social sciences
\citep{mccrea2014analysis, bohning2017capture}.
For example, management of conservation is facilitated by data on population numbers of  endangered species
\citep{perez2014estimating}, and details about increased
numbers of unskilled immigrants  may be important for income
maintenance programs because of the resulting increased costs  \citep{borjas1994economics}.
To estimate  abundance,
capture--recapture experiments constitute a widely used
sampling technique for data collection \citep{otis1978statistical}.
These experiments are based on a series of occasions or are conducted over a period of time,
during which individuals are captured,
marked or noted, and then released back into the population.
The collected data usually exhibit 
two features: individual heterogeneity and one-inflation
{\citep{chao2001overview, bohning2017capture, godwin2017estimation}.}

Individual heterogeneity has been
thought of as a key factor affecting the probability of being captured.
Ignoring  heterogeneity may
lead to severely biased estimates \citep{chao2001overview}.
A number of nonparametric abundance estimators, which account for unobserved heterogeneity,
have been proposed, such as the jackknife estimator \citep{burnham1978estimation}, the estimator of
\cite{zelterman1988robust}, and
the lower  bound estimator of \cite{chao1987estimating}. 
As individual covariates are usually available in practice,
 heterogeneity among capture probabilities could be modeled by
parametric regression models on covariates,
e.g., binomial regression models for discrete-time capture--recapture data
and Poisson regression models for continuous-time capture--recapture data
\citep{chao2001overview, van2003point}.
In the presence of individual covariates, researchers have developed
partial likelihood, conditional likelihood (CL), and
semiparametric empirical likelihood (EL) methods
to make inference for abundance;
see \cite{huggins2011review} and \cite{liu2017maximum, liu2018full}.
Maximum EL estimators of  abundance usually
have smaller mean square errors than  CL-based estimators
and  EL ratio confidence intervals (CIs)
have more accurate coverage probabilities than  CL-based CIs.

%

One-inflation problems have recently
received much attention in the context of capture--recapture models or
zero-truncated count models.
\cite{godwin2017estimation} first noticed that
the number of captures usually exhibits a preponderance of ``1''-counts.
Ignoring excessive 1s may
result in upwards biased estimates of population sizes.
To account for the excess 1s,
\cite{godwin2017estimation}
proposed using a one-inflated zero-truncated Poisson distribution
(see the model \eqref{eq:cap-prob-inflate-positive-con} in Section \ref{s:ext})
and a  zero-truncated one-inflated Poisson distribution
(see the model \eqref{eq:cap-prob-zero-trun} in Section \ref{s:mod-data})
to model the probability mass function of the number of captures,
and they suggested a score test for the one-inflation parameter.
More complicated count models
have been  investigated in the literature,
such as the one-inflated zero-truncated negative binomial model
\citep{godwin2017one, inan2018one}, the
one-inflated zero-truncated Poisson mixture model
\citep{godwin2019one},
and the one-inflated zero-truncated geometric model
\citep{bohning2019identity, bohning2021population}.
The zero-truncated one-inflated geometric model was also investigated in
Chapter 14 of \cite{bohning2017capture}.
Recently,
\cite{bohning2020general}
considered general count models allowing both inflation and deflation.
Even though these models account for unobserved heterogeneity to some extent,
quite a few  researchers have resorted to regression models on individual covariates
to characterize the heterogeneity among capture probabilities.
As noted in the discussion sections of \cite{bohning2019identity}
and \cite{bohning2021population},
the inclusion of covariates would be helpful to improve the fit
of  models and increase the likelihood of valid abundance predictions.

In this paper, we simultaneously consider the observed heterogeneity
and one-inflation problems in the estimation of abundance from capture--recapture data.
Under the one-inflated zero-truncated Poisson regression model,
\cite{godwin2017estimation} proposed a Horvitz--Thompson-type estimator for the abundance
and a CL-based score test for the one-inflation parameter.
Under the one-inflated zero-truncated negative binomial regression model,
\cite{godwin2017one} also proposed a Horvitz--Thompson-type estimator
and CL-based likelihood ratio tests for the one-inflation and unobserved heterogeneity.
\cite{inan2018one} further
investigated the computation problem
in terms of a mean-parameterized negative binomial distribution.
However, the aforementioned methods  have at least two limitations.
First,  Horvitz--Thompson-type estimators  for abundance
may be unstable if  some detection probabilities are very close to zero.
Meanwhile,  variance estimations of these estimators have not been  discussed.
Second, these methods are inapplicable to
the one-inflated zero-truncated binomial regression model
and to zero-truncated one-inflated regression models.

We systematically study  inference problems concerning abundance
under  ``zero-truncated one-inflated''  and  ``one-inflated zero-truncated''
binomial and Poisson regression models,
with  available covariates incorporated   in each model to explain the observed heterogeneity.
Our contribution can be summarized as follows.
First, for each considered model, we propose a new score test
to test the existence of one-inflation.
The score test statistic is derived from a full likelihood and
is expected to be locally more powerful.
Second, we extend the EL method for  usual capture--recapture data
 to   one-inflated capture--recapture data, with
the resulting EL method inheriting the advantages over the CL method.
Third, to maximize the EL function,
we develop an efficient  expectation-maximization (EM) algorithm,
rather than using  existing optimization functions such as the \texttt{R} function
\texttt{nlminb()} adopted by \cite{liu2017maximum, liu2018full}.
The EM algorithm has an appealing convergence property and is free of  Lagrange multipliers.
Further, our numerical experience indicates that
the estimation results from the EM algorithm
are more stable and reliable than those derived from  existing optimization functions in \texttt{R}.

The rest of the paper is organized as follows.
In Section \ref{s:mod},
we introduce the zero-truncated  one-inflated regression models,
and we present the semiparametric EL approach to abundance estimation
and  the score test for the one-inflation parameter.
The EM algorithm is also discussed for  numerical implementation.
In Section \ref{s:ext},
the proposed method and algorithm are  extended to
one-inflated zero-truncated regression models.
In Section \ref{s:sim},
several simulation studies are conducted to examine the
finite-sample performance of the proposed methods.
In Section \ref{s:dat}, we apply the proposed approaches
to two real-life data sets.
Section \ref{s:dis} concludes with a discussion.
For convenience of presentation,
all proofs are given in the online Supporting Information.

\section{Semiparametric empirical likelihood inference under zero-truncated one-inflated regression models}
\label{s:mod}

\subsection{Model and data}
\label{s:mod-data}

Consider a closed population composed of $N$ individuals.
For a generic individual,
we denote by $\bX$ and $Y$
the 
individual covariate
and the number of times of being captured,
respectively.
Given $\bX = \bx$,
we assume that $Y$ follows a one-inflated count distribution
with the conditional probability mass function
\ba\label{eq:cap-prob}
h(y, \bx; \bbeta, w)   =
w f(y, \bx;\bbeta)  +
(1-w)  \cdot I(y=1).
\ea
Here $0\leq1-w< 1$ is the proportion of one-inflation
{among the individuals in the population} and
$f(y, \bx;\bbeta)$ is the probability mass function
without one-inflation. 
  Under the model \eqref{eq:cap-prob},
the conditional distribution of $Y = y$
for an individual whose covariate is $\bx$
and who has been captured at least once is
\ba\label{eq:cap-prob-zero-trun}
\pr(Y = y \mid \bX = \bx, Y>0)
=
\frac{ 1-w   }{1 - f(0, \bx;\bbeta)}\cdot I(y=1)	
+
\frac{w  f(y, \bx;\bbeta)}{1 - f(0, \bx;\bbeta)} \cdot I(y\geq 1).
\ea
In \eqref{eq:cap-prob-zero-trun},
the count process has one-inflation first and then zero-truncation next.
Hence, the model \eqref{eq:cap-prob-zero-trun} is called the
``zero-truncated one-inflated  regression model."

For discrete-time capture--recapture data,
$f(y, \bx;\bbeta)$ is usually chosen to be
a binomial distribution.
That is, for $y= 0, \ldots,K,$
\bas
f(y, \bx;\bbeta) = {K \choose y} \{g(\bx; \bbeta)\}^{y} \{1 - g(\bx; \bbeta)\}^{K- y},\quad
\mbox{with}\quad
g(\bx; \bbeta) = \frac{\exp(\bbeta^\T \bx)}{1 + \exp(\bbeta^\T \bx)},
\eas
where $K$ is the number of capture occasions.
For continuous-time capture--recapture data,
$f(y, \bx;\bbeta)$
 can be chosen to be
a Poisson distribution
with  probability mass function  
\bas
f(y, \bx; \bbeta) =
\frac{\{\lambda(\bx; \bbeta)\}^y}{y!}
\exp\{ - \lambda(\bx; \bbeta) \},\quad
\lambda(\bx; \bbeta) =
\exp(\bbeta^\T \bx),
\quad y= 0, 1,\ldots.
\eas

Suppose $\{(\bX_i, Y_i):i=1, \dots, N\}$ are independent and
identically distributed copies of $(\bX, Y)$ for
$N$ individuals in the population under study.
Without loss of generality,
we assume that the first $n$ individuals have been captured at least once
and the last $N-n$ individuals have not been captured at all.
For clarity, we denote the observations by
$\{(\bX_i = \bx_i,Y_i=y_i): i = 1,\dots,n\}$.
Our primary goal is to estimate the abundance $N$.
Under  a zero-truncated one-inflated  regression model,
we may wonder whether  one-inflation does in fact exist.
This leads to our second goal: testing  $H_0: w = 1$
under the  model \eqref{eq:cap-prob} or
\eqref{eq:cap-prob-zero-trun}
based on the observed data.
In the next subsection, we develop a semiparametric EL and
take it as the foundation of our subsequent statistical inference.

\subsection{Abundance estimation}
\label{sec:est-abun}
Based on the observations,
the full likelihood is
\bas
\pr(n) \times\prod_{i=1}^n \pr(\bX_i=\bx_i \mid Y_i>0)
\times\prod_{i=1}^n\pr(Y_i=y_i \mid \bX_i=\bx_i, Y_i>0).
\eas
Let $\alpha = \pr(Y=0)$ denote the probability of an individual  never being  captured.
The number of individuals captured, $n$,
follows a binomial distribution $\mathrm{Bi}(N, 1 - \alpha)$.
Therefore, the first term in the full likelihood is $\pr(n) = {N\choose n}(1-\alpha)^{n}\alpha^{N-n}$.
For the second and third terms, it follows from the Bayes' rule that
\bas
\prod_{i=1}^n\pr(\bX_i=\bx_i \mid Y_i>0) &=&
\prod_{i=1}^n
\frac{\pr( Y_i>0 \mid \bX_i=\bx_i )\pr(\bX_i=\bx_i)}{1- \alpha},
\eas
and
\bas
\prod_{i=1}^n\pr(Y_i=y_i \mid \bX_i=\bx_i, Y_i >0) &=&
\prod_{i=1}^n
\frac{\pr( Y_i = y_i \mid \bX_i=\bx_i )}{\pr( Y_i>0 \mid \bX_i=\bx_i )}.
\eas
Combining the above equations, we have the full likelihood
\bas
{N\choose n}\alpha^{N-n} \times
\prod_{i=1}^n \pr( Y_i = y_i \mid \bX_i=\bx_i )
\times \prod_{i=1}^n \pr(\bX_i=\bx_i).
\eas

We use EL
\citep{owen1988empirical,
owen1990empirical} to handle the nonparametric marginal distribution $F(\bx)$ of $\bX$.
Following the principle of EL, we model $F(\bx)$ by a discrete distribution
$ \sum_{i=1}^n p_i I(\bx_i \leq \bx)$,
where $p_i\geq 0$ ($i=1,\dots,n$) and
$\sum_{i=1}^n p_i=1$.
After substituting $p_i = \pr(\bX_i=\bx_i)$ into the likelihood and
taking the  logarithm, we get the log-EL
\ba\label{eq:em-like}
\widetilde \ell(N, \bbeta, w, \alpha, \{p_i\}) =
\log{N\choose n} + (N-n) \log(\alpha) +
\sum_{i=1}^n \log\{h(y_i, \bx_i;\bbeta, w)\} +
\sum_{i=1}^n\log(p_i).
\ea

Since $\alpha = \e\{\pr(Y=0 \mid \bX)\}$,
it follows from the model \eqref{eq:cap-prob} that
a feasible constraint on $F$ or the $p_i$ is
$
\sum_{i=1}^n\{ wf(0, \bx_i;\bbeta) - \alpha\} p_i = 0.
$
Using the method of Lagrange multipliers,
the maximum of \eqref{eq:em-like} is achieved at
$
p_i = n^{-1} [ 1 + \xi\{ wf(0,\bx_i;\bbeta)- \alpha\}]^{-1},\quad
i = 1,\dots,n.
$
Substituting this equation into \eqref{eq:em-like}
gives the profile log-EL of $(N, \bbeta, w, \alpha)$:
\begin{align}\label{eq:profile-EL}
\ell(N, \bbeta, w, \alpha) ={} &\log {N\choose n} + (N-n) \log(\alpha) +
\sum_{i=1}^n \log\{h(y_i, \bx_i;\bbeta, w)\}\nonumber\\
&- \sum_{i=1}^n
\log[1 + \xi\{ wf(0,\bx_i;\bbeta)- \alpha\}] - n\log(n),
  \end{align}
where $\xi = \xi(\bbeta, w, \alpha)$ satisfies
$$
\sum_{i=1}^n\frac{ wf(0, \bx_i;\bbeta) - \alpha}
{1 + \xi\{ wf(0,\bx_i;\bbeta)- \alpha\}} = 0.
$$
\begin{remark}
Under   $H_0: w=1$,
the model \eqref{eq:cap-prob} or
\eqref{eq:cap-prob-zero-trun}
degenerates to the case in which
there is no one-inflation.
In that situation,
the semiparametric EL methods were first proposed by \cite{liu2017maximum} in
discrete-time capture--recapture studies
and then extended to continuous-time capture--recapture data by \cite{liu2018full}.
From this point of view, this paper can be seen as an extension of
previous works to  one-inflated capture--recapture data.
\end{remark}

Based on the profile log-EL,
we define the maximum EL estimator of $(N, \bbeta, w, \alpha)$ as
$
(\widehat N, \widehat\bbeta, \widehat w, \widehat\alpha) =
\arg\max_{(N, \bbeta, w, \alpha)}
\{\ell(N, \bbeta, w, \alpha)\}
$
and the EL ratio function of $N$ as
$
R(N) =2\{\ell(\widehat N, \widehat\bbeta, \widehat w, \widehat\alpha) -
 \max_{(\bbeta, w, \alpha)}\\\ell(N, \bbeta, w, \alpha)\}.
$
Before discussing the asymptotic behaviors of
$(\widehat N, \widehat\bbeta, \widehat w, \widehat\alpha)$
and the EL ratio statistic,
we present an efficient numerical algorithm for
implementing the EL methods.


\subsection{EM algorithm}\label{sec:em}
A key step in the implementation of  the semiparametric EL method
is to maximize the log-EL function 
\eqref{eq:em-like} or
the profile version \eqref{eq:profile-EL}.
Similar to \cite{liu2017maximum} and \cite{liu2018full},
one can maximize $\ell(N, \bbeta, w, \alpha)$ by
using existing optimization functions such as the \texttt{R} function \texttt{nlminb()}.
However, there is no theoretical guarantee that these functions work
 in the current setup.
Further, when implementing the CL method in the absence of covariates,
\cite{godwin2017estimation} noticed that
their EM algorithm gives estimates with
slightly less bias  and smaller mean squared error,
compared with those from  packaged numerical methods.
In this subsection,
we develop an efficient EM algorithm for the numerical implementation of the proposed methods.

We first fix $N$,  and we would like to maximize  $\ell(N, \bbeta, w, \alpha)$
or $\widetilde \ell(N, \bbeta, w, \alpha, \{p_i\})$ with the given $N$.
In this situation, there are two types of missing information
in the context of the one-inflated capture--recapture data:
(1) the covariates are missing if individuals are not captured at all;
(2) it is not known whether the event that an individual was captured exactly once
is attributed to the one-inflation or not.
To describe the second type of missing data,
we introduce a latent variable $Z_i$ for individual $i$
that follows $\mathrm{Bi}(1, w)$ and is independent of $\bX_i$
($i=1,\ldots, N$). 
Given $\bX_i= \bx_i$,
$Y_i$ is assumed to have a probability mass function $f(y, \bx_i; \bbeta)$
for $Z_i = 1$ and follow a degenerate distribution at one
for $Z_i = 0$.
By this assumption, $Z_i=1$
if the observed value of $Y_i$ is not equal to one.

Without loss of generality,
we assume that $y_1=\dots=y_m=1$ and $y_i>1$ for $i =m+1,\dots,n$.
Given $N$, we let $\bO$ and $\bO^*$ denote the observed data and missing data, respectively.
Then,
the complete data can be written as
$\bO\cup \bO^*$:
\begin{align*}
{\bO} = {}&\{(Y_i=y_i, \bX_i = \bx_i): i = 1,\dots,n\}
\cup\{Y_i=0: i = n+1,\dots,N\}
\\
&\cup\{Z_i=1: i = m+1,\dots,N\},\\
{\bO^*} ={}& \{ \bX_j = \bx_j^*, j = n+1,\dots, N\}
\cup\{Z_i=z_i^*: i = 1,\dots,m\},
\end{align*}
where 
$\bx_{j}^*$s 
stand for the unobserved covariates of
the individuals never captured
and $z_i^*$s 
stand for
the latent variables indicating 
whether $y_i$ arises from $f(y, \bx_i;\bbeta)$ or not.
Both the $\bx_{j}^*$ and the $z_{i}^*$ serve in the role of missing data in the framework of the EM algorithm
\citep{dempster1977maximum}.
Based on $\bf O\cup O^*$, the complete-data likelihood is
\bas
&&
\prod_{i = 1}^m
\{\pr(Y_i=y_i \mid \bX_i = \bx_i, Z_i=z_i^*)\pr( Z_i=z_i^*) \pr(\bX_i= \bx_i) \}\\
&&
\times
\prod_{i = m+1}^n
\{\pr(Y_i=y_i \mid \bX_i = \bx_i, Z_i=1)\pr( Z_i=1) \pr(\bX_i= \bx_i) \}\\
&&
\times
\prod_{j = n+1}^N
\{\pr(Y_j=0 \mid \bX_j = \bx_j^*, Z_j=1) \pr( Z_j=1) \pr(\bX_j = \bx_j^*)\}.
\eas

Let $\btheta = (\bbeta, w, \alpha,
\{p_i\})$.
Recall that the EL method models $F(\bx)$ as
$F(\bx) = \sum_{i=1}^n p_i I(\bx_i \leq \bx)$.
That is, $ \bx_j^*$ is equal  to one of the $\bx_i$.
Hence,
the complete-data log-likelihood of $\btheta$ can be written as
\bas
\ell_c(\btheta)&=&
\sum_{i = 1}^m [
I(z_i^*=0)\log(1-w)+ I(z_i^*=1)  \log\{wf(y_i, \bx_i;\bbeta)\} ]+
\sum_{i = 1}^n \log(p_i)
\\
&&+
\sum_{i = m+1}^n\log\{w f(y_i, \bx_i;\bbeta)\} +
\sum_{i = 1}^n\sum_{j = n+1}^N [ I(\bx_j^* = \bx_i)
\log\{w f(0, \bx_i;\bbeta) p_i\} ].
\eas

The core of an EM algorithm is the EM iteration,
which consists of  an E-step and an M-step. 
These steps are iterated until convergence.
Let $\btheta^{(0)}$ be the initial value of $\btheta$.
For $r = 1, 2, \dots$, we denote by
$\btheta^{(r)} = (\bbeta^{(r)}, w^{(r)},
\alpha^{(r)},\{p_i^{(r)}\})$
 the value of $\btheta$
after $r$ rounds of EM iterations.
In the E-step of the $(r+1)$th iteration,
we need to calculate
$Q(\btheta|\btheta^{(r)}) =
\e\{\ell_c(\btheta)|{\bO}, \btheta^{(r)}\}$.
Given $\bO$ and
using $\btheta^{(r)}$ for $\btheta$,
the conditional expectation of
$I(z_i^*=1)$ for $i = 1,\dots, m$ is
\bas
v_i^{(r)}= \e\{I(z_i^* = 1) \mid {\bO}, \btheta^{(r)}\} &=&
\frac{\pr(Z_i = 1, Y_i = 1 \mid \bX_i = \bx_i,  \btheta = \btheta^{(r)} )}
{\pr(Y_i = 1 \mid  \bX_i = \bx_i,  \btheta = \btheta^{(r)} )}
\\
&=&
\frac{w^{(r)} f(1, \bx_i; \bbeta^{(r)})}{1-w^{(r)}+w^{(r)} f(1, \bx_i; \bbeta^{(r)})} ,
\eas
and the conditional expectation of
$I(\bx_j^*=\bx_i)$ for $j = n+1,\dots, N$
and $i = 1, \dots, n$ is
\bas
\e\{I(\bx_j^* = \bx_i) \mid {\bO}, \btheta^{(r)}\} =
\pr(\bX_j = \bx_i \mid Y_j = 0,  \btheta = \btheta^{(r)})
= \frac{ w^{(r)} f(0, \bx_i; \bbeta^{(r)}) p_i^{(r)} }{ \alpha^{(r)}},
\eas
where $ \alpha^{(r)} =  w^{(r)}
\sum_{i = 1}^n f(0, \bx_i; \bbeta^{(r)}) p_i^{(r)}$.
Thus, the conditional expectation of the complete-data log-likelihood is
\bas\label{eq:cdll}
Q(\btheta|\btheta^{(r)}) &=& \ell_1(w) + {\ell_2(\bbeta)} + \ell_3(\{p_i\}),
\eas
where
\bas
 \ell_1(w)&=&
\sum_{i = 1}^n (1-v_i^{(r)}) \log(1-w)
+ \sum_{i = 1}^n ( v_i^{(r)}+ u_i^{(r)}) \log(w),
\\
\ell_2(\bbeta)&=&
\sum_{i=1}^n [
v_i^{(r)} \log\{f(y_i, \bx_i;\bbeta)\} + u_i^{(r)} \log\{f(0, \bx_i;\bbeta)\}
],
\\
\ell_3(\{p_i\})&=&
\sum_{i = 1}^n (u_i^{(r)} + 1) \log(p_i),
\eas
with  $u_i^{(r)} = (N - n)w^{(r)} f(0, \bx_i; \bbeta^{(r)}) p_i^{(r)}/ \alpha^{(r)}$
for $i=1,\dots,n$,
and $v^{(r)}_i=1$ for $i=m+1,\dots,n$.

In the M-step of the $(r+1)$th iteration,
we update $\btheta$ by $\btheta^{(r+1)}$, which
maximizes $Q(\btheta|\btheta^{(r)})$
with respect to $\btheta$ under the constraints
$$
p_i \geq 0,\; i = 1, \dots, n,\quad
\sum_{i=1}^n p_i=1,\quad
\sum_{i=1}^n\{ wf(0, \bx_i;\bbeta) - \alpha\} p_i = 0.
$$
Since the three parts in $Q(\btheta|\btheta^{(r)})$
are functions of $w$, $\bbeta$, and $p_i$s, respectively,
the M-step could be easily implemented through the following steps.

\begin{description}

\item[\bf Step 1]
Maximizing $\ell_1(w)$ gives the updated value of $w$:
$
w^{(r+1)} =  \sum_{i = 1}^n ( v_i^{(r)}+ u_i^{(r)})/\sum_{i = 1}^n (1 + u_i^{(r)}).
$

\item[\bf Step 2]
Update $\bbeta$ to $\bbeta^{(r+1)}$ by maximizing
$\ell_2(\bbeta)$. 
This can be implemented 
by fitting a generalized linear regression model
to the data $\{(y_1, \bx_1),\dots, (y_n, \bx_n), (0, \bx_1),\dots, (0, \bx_n)\}$,
with the weights being $(v_1^{(r)},\dots,v_n^{(r)}, u_1^{(r)},\dots, u_n^{(r)})$.
For example, it can be easily done by invoking the {\tt R} function
{\tt glm()} with the link function being ``logit''  for the binomial case and
``log'' for the Poisson case.

\item[\bf Step 3]
Update the $p_i$ to  the maximizer of
$\ell_3(\{p_i\})$, namely,
$
p_i^{(r+1)} =  ( u_i^{(r)} + 1)/  \sum_{i=1}^n (u_i^{(r)} + 1), \quad
i=1,\dots,n.
$

\item[\bf Step 4]
Update $\alpha$ to
$\alpha^{(r+1)} = w^{(r+1)}\sum_{i = 1}^n f(0, \bx_i; \bbeta^{(r+1)})p_i^{(r+1)}$.

\end{description}

We make several comments on the above EM algorithm.
First,
following the proof in \cite{dempster1977maximum},
we can show that the log-EL
$\widetilde \ell(N, \bbeta, w, \alpha, \{p_i\})$ in  \eqref{eq:em-like}  does not
decrease after each iteration. Further, note that
$\widetilde\ell(N, \bbeta, w, \alpha, \{p_i\})\leq 0$.
Thus, the sequence $(\bbeta^{(r)}, w^{(r)}, \alpha^{(r)}, \{p_i^{(r)}\})$
eventually converges to a stationary point of
$\widetilde\ell(N, \bbeta, w, \alpha, \{p_i\})$ for given $N$.
Or, equivalently,
the sequence $(\bbeta^{(r)}, w^{(r)}, \alpha^{(r)})$
converges to a stationary point of
$\ell(N, \bbeta, w, \alpha)$ for given $N$.
Second, in the M-step, the updated values of unknown parameters either have closed forms or can be easily obtained using existing \texttt{R} functions.
This makes the EM algorithm very stable and flexible.
Third, we stop the algorithm when the increment in the log-EL \eqref{eq:em-like} after an iteration is no greater than,
say, $10^{-5}$.

We now consider the case where $N$ is unknown and discuss the
numerical calculation of 
$(\widehat N, \widehat \bbeta, \widehat w, \widehat \alpha)$.
The  above EM algorithm can be easily adapted for this situation.
We need two modifications.
First,  we set $u_i^{(r)}$ to
$(N^{(r)} - n)w^{(r)} f(0, \bx_i; \bbeta^{(r)}) p_i^{(r)}/
\alpha^{(r)}$ in the $(r+1)$th iteration.
Second, we  add a maximization step to update $N$
after {\bf Step 4} as follows.

\begin{description}
\item[\bf Step 4$'$]
Update $N$ to $N^{(r+1)}$ by maximizing
$\log{N \choose n} + (N - n) \log(\alpha^{(r+1)})$.
This step can be implemented by
using the  existing {\texttt R} function {\tt optimize()}.
\end{description}

The following theorem summarizes the properties of
the EM algorithm discussed above.
\begin{theorem}
\label{thm0}
With the EM algorithms described above, we have for 
$r=0,1,\dots,$
\begin{enumerate}
\item[(a)]
$
\widetilde\ell(N, \bbeta^{(r+1)}, w^{(r+1)}, \alpha^{(r+1)}, \{p_i^{(r+1)}\})
\geq\widetilde\ell(N, \bbeta^{(r)}, w^{(r)}, \alpha^{(r)}, \{p_i^{(r)}\})
$
when $N$ is fixed at each iteration;
\item[(b)]
$
\widetilde\ell(N^{(r+1)}, \bbeta^{(r+1)}, w^{(r+1)}, \alpha^{(r+1)}, \{p_i^{(r+1)}\})
\geq\widetilde\ell(N^{(r)}, \bbeta^{(r)}, w^{(r)}, \alpha^{(r)}, \{p_i^{(r)}\})
$
when $N$ is updated at each iteration.
\end{enumerate}
\end{theorem}

%
%

\subsection{Asymptotic property}\label{sec:asy-pro}
In this subsection, we present the asymptotic properties of
the maximum EL estimator
$(\widehat N, \widehat \bbeta, \widehat \alpha, \widehat w)$ and
the EL ratio statistic $R(N_0)$.
Before that, we need some notation.
Let $(N_0, \bbeta_0, w_0, \alpha_0)$ be the true value of
$(N, \bbeta, w, \alpha)$ with $0<\alpha_0<1$ and $0<w_0\leq 1$.
We use $e_f(\bx;\bbeta)$ and $v_f(\bx;\bbeta)$ to denote the conditional expectation
and variance, respectively,  of $Y$  conditional on $\bX=\bx$.
For any vector or matrix $\bA$, we use $\bA^{\otimes2}$
to denote $\bA \bA^\T$.
Define $\varphi = \e\{1 - w_0f(0, \bX; \bbeta_0)\}^{-1}$
and
$$
\bW =
\begin{pmat}[{....}]
-\rV_{11}& \bzero^\T& 0 & - \rV_{14}\cr
\bzero &- \bV_{22}+\bV_{25}\rV_{55}^{-1}\bV_{52}
& - \bV_{23} + \bV_{25}\rV_{55}^{-1}\rV_{53}
& - \bV_{24} + \bV_{25}\rV_{55}^{-1}\rV_{54}\cr
0  &- \bV_{32}+ \rV_{35}\rV_{55}^{-1} \bV_{52}
& -  \rV_{33} +  \rV_{35}\rV_{55}^{-1} \rV_{53}
& -  \rV_{34} + \rV_{35}\rV_{55}^{-1} \rV_{54}\cr
-\rV_{41}  &- \bV_{42}+ \rV_{45}\rV_{55}^{-1} \bV_{52}
& -  \rV_{43} +  \rV_{45}\rV_{55}^{-1} \rV_{53}
& -  \rV_{44} + \rV_{45}\rV_{55}^{-1} \rV_{54}\cr
\end{pmat},
$$
where $\bzero$ is the zero vector,
$\rV_{11} = 1- \alpha_{0}^{-1}$, $\rV_{14} = \rV_{41} = \alpha_{0}^{-1}$, and
\bas
\bV_{22} &=& w_0
\e \!\left[
	\frac{ f(0, \bX; \bbeta_0)  \{e_f(\bX;\bbeta_0)\}^2 \bX^{\otimes2}}
	{1 - w_0 f(0, \bX; \bbeta_0)} - v_f(\bX;\bbeta_0) \bX^{\otimes2}\right.
\\&&\qquad
\left.+
	\frac{ (1 - w_0) f(1, \bX; \bbeta_0)  \{1 - e_f(\bX;\bbeta_0)\}^2 \bX^{\otimes2}}
	{1 - w_0 + w_0 f(1, \bX; \bbeta_0)} \right],
\\
\bV_{23} &=& \bV_{32}^\T = \e \!\left[
	\frac{f(1,\bX;\bbeta_0) \{1- e_f(\bX;\bbeta_0)\} \bX}{1 - w_0 + w_0 f(1,\bX;\bbeta_0)}
	- \frac{ f(0,\bX;\bbeta_0)e_f(\bX;\bbeta_0) \bX}{ 1 - w_0 f(0, \bX;\bbeta_0) }
	\right],
\\
\rV_{33} &=&
	\e\!\left[ -
	\frac{\{1 - f(1, \bX;\bbeta_0)\}^2}{1 - w_0 + w_0 f(1, \bX;\bbeta_0)} -
	\frac{1 - f(0, \bX;\bbeta_0) -  f(1, \bX;\bbeta_0)}{w_0} +
	\frac{\{ f(0, \bX;\bbeta_0) \}^2}{1 - w_0 f(0, \bX; \bbeta_0)}
	\right],
\\
\bV_{24} &=& \bV_{42}^\T =
	 \e\! \left\{
	\frac{ w_0 f(0, \bX; \bbeta_0) e_f(\bX;\bbeta_0) \bX}{ 1 - w_0 f(0, \bX; \bbeta_0) }
	\right\}, \quad
\bV_{25} = \bV_{52}^\T =	(1-\alpha_0)^2 \bV_{24},
\\[6pt]
\rV_{34} &=& \rV_{43}  = -
	\e\! \left\{
	\frac{ f(0, \bX; \bbeta_0) }{ 1 - w_0 f(0, \bX; \bbeta_0)  }
	\right\}, \quad
\rV_{35} = \rV_{53}  =
	(1 - \alpha_0)^2 \rV_{34},
\\[6pt]
\rV_{44} &=&\varphi- \alpha_0^{-1},
\quad
\rV_{45} = \rV_{54}  =
	(1 - \alpha_0)^2 \varphi,\quad
\rV_{55} =
	(1 - \alpha_0)^4\varphi - (1 - \alpha_0)^3.
\eas

In the following theorem, we discuss
the asymptotic properties
of  $(\widehat N, \widehat \bbeta, \widehat w, \widehat \alpha)$
and $R(N_0)$ when
there exists one-inflation under the model \eqref{eq:cap-prob} or
\eqref{eq:cap-prob-zero-trun}.

\begin{theorem}\label{the-asy}
Suppose that $0<w_0<1$ and  $\bW$ is positive definite.
As $N_0\to \infty,$ we have
\begin{enumerate}
\item[(a)]
$\sqrt{N_0} \{ (\widehat N-N_0)/N_0, (\widehat\bbeta-\bbeta_0)^\T,
\widehat w - w_0, \widehat\alpha - \alpha_0\}^\T
\convergeto {\rm N}(\bzero,\bW^{-1}),$
where $\convergeto$ denotes convergence in distribution;
\item[(b)]
  $\sqrt{N_0} (\widehat N / N_0 - 1) \convergeto {\rm N}(0, \sigma^2),$ where
$$
\sigma^2 = \varphi - 1 -
\begin{pmat}[{..}]
\bV_{42}&\rV_{43}\cr
\end{pmat}
\begin{pmat}[{..}]
\bV_{22}&\bV_{23}\cr
\bV_{32}&\rV_{33}\cr
\end{pmat}^{-1}
\begin{pmat}[{..}]
\bV_{24}\cr
\rV_{34}\cr
\end{pmat};
$$
\item[(c)]
$R(N_0)\convergeto \chi^2_{1},$ where $\chi^2_{1}$ stands for the chi-square distribution
with one degree of freedom.
\end{enumerate}
\end{theorem}

In part (b) of Theorem \ref{the-asy},
$\sigma^2$ is usually unknown.
To construct a consistent estimate for $\sigma^2$, we note that the
$\rV_{ij}$ and $\varphi$ can be expressed as
$\e\{J(\bX; \bbeta_0, w_0, \alpha_0)\}$
for some function $J$,
for which a consistent estimator is 
$
(\widehat N)^{-1} \sum_{i=1}^{n}
 J(\bx_i; \widehat \bbeta, \widehat w, \widehat \alpha)/\{ 1 - \widehat w f(0, \bx_i; \widehat \bbeta)\}.
$
A consistent estimator  of $\sigma^2$ can  therefore be obtained
by replacing the $\rV_{ij}$ and $\varphi$ by their respective estimators.
We denote this estimator by $\widehat\sigma^2$.

With parts (b) and (c) in Theorem \ref{the-asy},
we can construct  an EL ratio CI
and a Wald-type CI of $N$ as
$$
\mathcal{I}_1 = \{N: R(N) \leq \chi^2_1(1-a)\},\quad
\mathcal{I}_2 = \{N: (\widehat N - N)^2/(\widehat N \widehat\sigma^2) \leq
\chi^2_{1}(1-a)\},
$$
where $\chi^2_1(1-a)$ is the $(1-a)$th quantile of
the $\chi_1^2$ distribution.
Even though both $\mathcal{I}_1$ and $\mathcal{I}_2$
have  asymptotically correct coverage probabilities,
our simulation studies show that
the EL ratio CI $\mathcal{I}_1$ is usually
superior to the Wald-type CI $\mathcal{I}_2$.

{\subsection{Score test for the existence of one-inflation}
\label{section2.score}

One problem of practical and scientific interest is whether
 one-inflation exists.
Under the model \eqref{eq:cap-prob} or
\eqref{eq:cap-prob-zero-trun},
this is equivalent to testing whether $w=1$.
We  propose a score test
based on the semiparametric EL in \eqref{eq:profile-EL}
 for the null hypothesis $H_0: w=1$.

Let $(\widetilde N, \widetilde \bbeta, \widetilde \alpha)$ be the maximum EL estimator of $(N, \bbeta, \alpha)$ under $H_0$.
Taking the partial derivative of $\ell(N, \bbeta, w, \alpha)$
with respect to $w$ at $w=1$, and replacing the unknown parameters
by the corresponding maximum EL estimators under $H_0$,
 we obtain the score statistic
$
 \partial \ell(\widetilde N, \widetilde \bbeta, 1, \widetilde \alpha)/\partial w
=
U(\widetilde N,\widetilde \bbeta),
$
where  $U(N,\bbeta)=N - \sum_{i=1}^m \{f(1, \bx_i;  \bbeta)\}^{-1}$.
We refer to Section 2.3 of
the Supporting Information 
for a detailed derivation of this statistic.
Next, we study the  properties of $U(\widetilde N,\widetilde \bbeta)$
and use it to construct a score test for testing $H_0:w=1$.

Let
$
\sigma_u^2
= \rV_{66s} -
[1,\, \bV_s^\T, \,  0]
\bW_s^{-1}
[1,\, \bV_s^\T, \,  0]^\T,
$
where $
 \rV_{66s}=  \e
[\{f(1, \bX; \bbeta_0)\}^{-1}] - 1$,
$\bV_s = \e[\{1 - e_f(\bX; \bbeta_0)\}\bX]$,
and $\bW_s$ is defined in Equation (7) of
the Supporting Information.

\begin{theorem}\label{the:score-test}
Suppose  that $\rV_{66s} <\infty$ and the matrix $\bW_s$
is positive definite.
Then
\begin{enumerate}
\item[(a)] $\e\{ U(N_0,\bbeta_0)  \}\leq 0,$ with  equality holding if and only if $w_0=1;$
\item[(b)]
under the null hypothesis $H_0: w=1,$
as $N_0\to \infty,$
$
N_0^{-1/2}U (\widetilde N,\widetilde \bbeta) \convergeto N(0,\sigma_u^2).
$
\end{enumerate}
\end{theorem}

To construct a score test based on $U (\widetilde N,\widetilde \bbeta)$,
we need a consistent estimator for $\sigma_u^2$.
This can be achieved by  techniques similar to those in deriving $\widehat\sigma^2$ in Section \ref{sec:asy-pro}.
We denote the resulting consistent estimator of $\sigma_u^2$ by $\widetilde\sigma_u^2$.
With the results in Theorem \ref{the:score-test} and $\widetilde\sigma_u^2$,
our score test statistic for $H_0:w=1$ is defined as
$
S= U (\widetilde N,\widetilde \bbeta)/\{ \widetilde N^{1/2} \widetilde\sigma_u \},
$
which converges in distribution to $N(0,1)$ under $H_0$.
At  significance level $a$,
we reject the null hypothesis  $H_0: w=1$ when $S\leq Z_{a}$, where $Z_a$ is the $a$th quantile of $N(0,1)$.
}

\section{Extension to one-inflated zero-truncated regression models}
\label{s:ext}


The proposed semiparametric EL approach is so flexible that it is applicable
 not only  to zero-truncated one-inflated  regression models, but also to
one-inflated zero-truncated regression models.
Given $\bX = \bx$, a one-inflated zero-truncated regression model
assumes that  the conditional probability mass function of $Y$ is
\ba\label{eq:cap-prob-inflate-positive}
h_e(y, \bx; \bbeta, w )   =
\begin{cases}
f(0, \bx;\bbeta), 				&y=0,\\
(1-w )\{1 - f(0, \bx;\bbeta)\} +
w  f(1, \bx;\bbeta), 	&y=1,\\
w  f(y, \bx;\bbeta), 				&y > 1,
\end{cases}
\ea
where $f(y, \bx;\bbeta)$ is the probability mass function defined in
Section \ref{s:mod-data}
and $0\leq 1-w < 1$ describes the proportion of one- and zero-inflation
in the population under study.
Under the model \eqref{eq:cap-prob-inflate-positive},
the conditional distribution of $Y = y$
for an individual whose covariate is $\bx$
and who has been captured at least once is
\ba\label{eq:cap-prob-inflate-positive-con}
\pr(Y = y \mid \bX = \bx, Y>0)
 =
 (1-w ) \cdot I( y=1)
+
\dfrac{w  f(y, \bx;\bbeta)}{1 - f(0, \bx;\bbeta)}\cdot I(y\geq 1).
\ea
As discussed in \cite{godwin2017estimation},
the expression in  \eqref{eq:cap-prob-inflate-positive-con}
looks as if the count  process were first  truncated at zero and then
had one-inflation.
Hence, we call the model \eqref{eq:cap-prob-inflate-positive-con}
 a ``one-inflated zero-truncated regression model.''
Note that  $1-w$ also describes the proportion of individuals in the sample 
who learned avoidance ability at their first capture
and thus were not  captured subsequently.

With   arguments  similar to those in Section \ref{sec:est-abun},
we obtain the log-EL function, which has the same form
as \eqref{eq:em-like} with $h(y, \bx; \bbeta, w)$ replaced by $h_e(y, \bx; \bbeta, w )$.
Under the model \eqref{eq:cap-prob-inflate-positive} or
\eqref{eq:cap-prob-inflate-positive-con},
the profile log-EL becomes
\begin{align}\label{eq:profile-EL-inflate-positive}
\ell_e(N, \bbeta, w , \alpha) = {}&\log{N\choose n} + (N-n) \log(\alpha) +
\sum_{i=1}^n \log\{h_e(y_i, \bx_i;\bbeta, w )\}\nonumber
\\
& - \sum_{i=1}^n
\log[1 + \xi\{ f(0,\bx_i;\bbeta)- \alpha\}],
\end{align}
where $\xi = \xi(\bbeta, \alpha)$ satisfies
$$
\sum_{i=1}^n\frac{ f(0, \bx_i;\bbeta) - \alpha}
{1 + \xi\{ f(0,\bx_i;\bbeta)- \alpha\}} = 0.
$$
Based on \eqref{eq:profile-EL-inflate-positive},
we define the maximum EL estimator of $(N,\bbeta, w ,\alpha)$ as
$
(\widehat N_e, \widehat\bbeta_e, \widehat w _e, \widehat\alpha_e) =
\arg\max_{(N, \bbeta, w, \alpha)}  \\ \ell_e(N, \bbeta, w , \alpha)
$
and the EL ratio function of $N$ as
$
R_e(N) =2\{ \ell_e(\widehat N_e, \widehat\bbeta_e,
\widehat w _e, \widehat\alpha_e) -
 \max_{(\bbeta, w , \alpha)} \ell_e(N, \bbeta, w , \alpha) \}
.
$
With slight modifications,
the EM algorithm proposed in Section \ref{sec:em}
can  still be used to calculate the EL estimators.
See Section 3 of the Supporting Information for  details.

{Similar to Section \ref{section2.score}, we can construct a score test for $H_0:w =1$ based on \eqref{eq:profile-EL-inflate-positive}.
The difference between \eqref{eq:profile-EL} and
\eqref{eq:profile-EL-inflate-positive} lies in that
the constraint of the Lagrange multiplier $\xi$
in \eqref{eq:profile-EL} is
related to the one-inflation parameter $w$, but
the $\xi$ in \eqref{eq:profile-EL-inflate-positive} is not.
This results in a different score function for the null hypothesis
$H_0: w =1$:
$$
U_e(\bbeta):=\frac{\partial \ell_e( N,   \bbeta, 1,  \alpha)}{\partial w} = \sum_{i=1}^n \frac{\dot h_e(y; \bx_i;
\bbeta, 1)}
{h_e(y; \bx_i; \bbeta, 1)}=
n -
\sum_{i=1}^m
\frac{1 - f(0; \bx_i;  \bbeta)}
{f(1; \bx_i;  \bbeta)}.
$$
Since $\bbeta$ is unknown, we estimate it by its maximum EL estimator $\widetilde\bbeta$ under $H_0$
and obtain $U_e(\widetilde\bbeta)$.

To present some asymptotic properties of $U_e(\widetilde\bbeta)$, we define
\bas
\sigma_{ue}^2
&=&
\rV_{66e}
+
\begin{pmat}[{....}]
1& \bV_{se}^\T & 0\cr
\end{pmat}
\bW_s^{-1}
\begin{pmat}[{...}]
1\cr
\bV_{se}\cr
0\cr
\end{pmat} +
2\begin{pmat}[{....}]
1& \bV_{se}^\T & 0\cr
\end{pmat}
\bW_s^{-1}
\begin{pmat}[{.}]
0\cr
\bV_{26e}\cr
0\cr
\end{pmat},
\eas
where
$
\rV_{66e}=  \e
[\{1 - f(0, \bX;\bbeta_0)\}^2\{f(1, \bX; \bbeta_0)\}^{-1}] -
(1 - \alpha_0)$,
$\bV_{se}=
\e[\{1 - f(0, \bX;\bbeta_0) - e_f(\bX; \bbeta_0)\}\bX]$,
and
$$
\bV_{26e} = \e\!\left[\left\{
\frac{e_f(\bX; \bbeta_0)}{1 - f(0, \bX; \bbeta_0)} - 1 +
f(0, \bX; \bbeta_0) - f(0, \bX; \bbeta_0)e_f(\bX; \bbeta_0)
\right\}\bX\right].
$$


\begin{theorem}\label{the:score-test2}
Let 
$\bW_e$ and $\sigma_e^2$ be two quantities defined
in Section 1.2 of the Supporting Information. 
Suppose that the matrices $\bW_s$ and
$\bW_e$ are positive definite.
\begin{enumerate}
\item[(a)]
Assume  $0<w _0<1$.  As $N_0\to\infty,$
$\sqrt{N_0} \{ ( \widehat N_e-N_0)/N_0, (\widehat\bbeta_e-\bbeta_0)^\T, \widehat w _e - w_0, \widehat\alpha_e - \alpha_0\}^\T \convergeto {\rm N}(\bzero,\bW_e^{-1}),$
$\sqrt{N_0} (\widehat N_e / N_0 - 1) \convergeto {\rm N}(0, \sigma_e^2),$
 and
$R_e(N_0)\convergeto \chi^2_{1}$.
\item[(b)] $\e\{U_e(\bbeta_0)\}\leq 0 ,$ with  equality holding if and only if $w_0=1$.
\item [(c)] Under the null hypothesis $H_0: w=1,$
as $N_0\to \infty,$
$
N_0^{-1/2}U_e(\widetilde\bbeta) \convergeto N(0,\sigma_{ue}^2).
$
\end{enumerate}
\end{theorem}

Using the technique to obtain $\widehat\sigma^2$ described in Section \ref{sec:asy-pro},
we can similarly obtain  consistent estimators
(denoted by $\widehat\sigma_e^2$ and $\widetilde\sigma_{ue}^2$)
for $\sigma_e^2$ and $\sigma_{ue}^2$.
Based on part (a) of Theorem \ref{the:score-test2},
we define the EL ratio confidence interval of $N$ as
$
\mathcal{I}_{1e} = \{N: R_e(N) \leq \chi^2_1(1-a)\}
$
and the Wald-type confidence interval as 
$
\mathcal{I}_{2e} = \{N: (\widehat N _e - N)^2/(\widehat N _e\widehat\sigma^2_e) \leq
\chi^2_{1}(1-a)\}.
$
With  parts (b) and (c) of Theorem \ref{the:score-test2} and
$\widetilde\sigma_{ue}^2$,
we define our score test statistic for $H_0:w=1$  as
$
S_e= U_e(\widetilde \bbeta)/( \widetilde N^{1/2} \widetilde\sigma_{ue}).
$
As  $\widetilde\sigma_{ue}$ is consistent,
when $N_0$ is large,
$S_e$ converges in distribution to ${\rm N}(0,1)$ under $H_0$.
We reject the null hypothesis of $w=1$ when $S_e\leq Z_{a}$  at  significance level $a$.
}

\section{Simulation studies}\label{s:sim}

In this section, we use simulation studies to
illustrate the finite-sample performance
of the proposed EL methods. 
We focus on Poisson regression models  here.
More simulation results  under binomial
regression models are given in Section 4 of
the Supporting Information.

Let $\bX = (1, X)^\T$ denote the individual covariate,
where $X \sim{\rm N}(18, 5)$ mimics
the covariate distribution  in drug user data  in Section \ref{s:dat}.
Given $\bX = \bx$, we generate the number of captures
from two scenarios.

\begin{enumerate}
\item[A.]
$Y$ is generated from the one-inflated model \eqref{eq:cap-prob}, where
$f(y, \bx; \bbeta)$ is a
Poisson regression model with the true value of regression coefficients being
$\bbeta_0 = (-1.5, 0.1)^\T$. In this scenario,
the capture probability ranges from 74\% to 92\%.
\item[B.]
This is the same as Scenario A except that $Y$ is generated from
the model \eqref{eq:cap-prob-inflate-positive}
with $\bbeta_0 = (-2.1, 0.1)^\T$. In this scenario,
the capture probability is about 49\%.
\end{enumerate}

In each scenario, we set $N_0 = 50$, 100, and 500.
 Based on simulated  data sets,
we evaluate the performance of the EL-based score tests
and  compare the maximum EL estimators
and EL ratio confidence intervals with those from existing methods.

\subsection{Comparison of score tests for the existence of one-inflation}
\label{sec:elrt}
For the null hypothesis $H_0 : w = 1$,
two EL-based score tests ($S$ and $S_e$) were derived
in Sections \ref{section2.score} and \ref{s:ext}.
As discussed, these two score test statistics asymptotically follow
a standard normal distribution.
Below, we compare their finite-sample performance with
the CL-based score test under the model
\eqref{eq:cap-prob-inflate-positive} or
\eqref{eq:cap-prob-inflate-positive-con},
denoted by $S_c = U_e(\widetilde \bbeta_c)/(\widetilde N_{HT}^{1/2} \widetilde\sigma_{uc})$,
where $\widetilde\bbeta_c$ is the maximum CL estimator of $\bbeta$ and
$\widetilde N_{HT}$ is the Horvitz--Thompson-type estimator of $N$
proposed by \cite{godwin2017estimation},
and $\widetilde\sigma_{uc}^2$ is
the estimator of $\var\{U_e(\widetilde \bbeta_c)\}/\widetilde N_{HT}$.
At  significance level $a$,
we reject $H_0: w=1$ when $S_c \leq Z_a$
and  define the $p$-value of $S_c$ as
the value of the cumulative distribution
function of $\mathrm{N}(0,1)$ at the observed  value of $S_c$,
rather than as half of the tail probability that
$\chi^2_1$ is greater than the observed value of  $S_c^2$
given in the code of \cite{godwin2017estimation}.

  We first check if the limiting distribution
provides an accurate approximation to the finite-sample
distributions of $S$, $S_e$, and $S_c$.
For this purpose, we consider Scenarios A and B with $w_0 = 1$.
In Table \ref{tab:sim-size}, we present
the simulated type I error rates of $S$, $S_e$, and $S_c$
at  significance levels 1\%, 5\%, and 10\%
based on 50,000 repetitions.
Overall, the type I error rates of all three tests are much closer to
the significance levels
in Scenario A, where the capture probability is high.
In Scenario B
with $N_0 = 50$,
the type I error rates of $S_e$ and  $S_c$ are usually underestimated,
and those of $S$ are slightly inflated.
By contrast, they become very close to the nominal significance levels
as $N_0$ increases to 500.
In summary, the limiting distributions of both
test statistics provide a satisfactory approximation
to their finite-sample distributions under both scenarios
when $N_0$ is large.

\begin{table}[h]
    \centering
    \caption{Simulated type I error rates (\%) of $S$, $S_e$, and $S_c$
    at  significance levels of
    1\%, 5\%, and 10\%}
    \label{tab:sim-size}
\begin{threeparttable}
    \begin{tabular}{ccccccccccccc}
    \toprule
	&&
	\multicolumn{3}{c}{$N_0=50$}
	&\phantom{}&
	\multicolumn{3}{c}{$N_0=100$}
	&\phantom{}&
	\multicolumn{3}{c}{$N_0=500$}
	
	\\

	Scenario & Level
	& 1\%	& 5\% 	&10\%&\phantom{}
	& 1\%	& 5\% 	&10\%& \phantom{}
	& 1\%	& 5\% 	&10\%
	\\
    \midrule
A&$S$&
1.15    &5.36   &10.54   &&
1.15    &5.38	&10.31 &&
1.02 	&5.20 	&10.38 %
\\
&$S_e$&
0.79 	&4.13 	&8.43  &&
0.90 	&4.56 	&8.94  &&
0.92 	&4.80 	&9.75
		\\
&$S_c$&
0.60 	&3.33 	&6.91  &&
0.73	&3.87	&7.68  &&
0.84 	&4.38 	&9.05
	\\

B&$S$&
1.47    &6.40	&11.93 &&
1.34    &6.03	&11.46 &&
1.16 	&5.31 	&10.43
\\
&$S_e$&
0.64 	&3.48 	&7.37  &&
0.81 	&4.08 	&8.55  &&
0.98 	&4.65 	&9.30
		\\
&$S_c$&
0.58	&2.90	&6.14  &&
0.70 	&3.43 	&7.13  &&
0.86 	&4.19 	&8.41 	
	\\
    \bottomrule
    \end{tabular}
\end{threeparttable}
\end{table}

We now compare the powers of the three score tests
under alternative models.
In each scenario,
we choose seven values of $w_0$  ranging
from 0.3 to 0.9 in increments of 0.1. 
In Table \ref{tab:sim-power}, we
present  the simulated powers
at a significance level of 5\%
based on 2000 repetitions.
We can see that the powers of $  S_e$
are close to or slightly higher than those of $S_c$ in most cases.
This may be because they are derived from
the same model
and the scores of
these two tests have the same closed form.
Relatively, the CL-based score test $S_c$ is not quite stable,
because 7\% and 13\% of the estimates $\widetilde \sigma_u^2$ are
negative in Scenario B with $w_0=0.3$ and 0.4, respectively.
For the score test $S$, its powers are comparable to
those of the other two tests when $N_0$ is as large as 500.
An interesting observation is that
as $N_0$ decreases to 100 and further to 50,
the score test $S$ derived from the
model \eqref{eq:cap-prob} always
dominates the score tests $S$  and $S_c$ derived from the
model \eqref{eq:cap-prob-inflate-positive}
whenever the true data-generating model is
\eqref{eq:cap-prob} or \eqref{eq:cap-prob-inflate-positive}.
A possible reason is that the model \eqref{eq:cap-prob}
is more parsimonious than the model \eqref{eq:cap-prob-inflate-positive},
and thus the observed data contain more information about
the former model when the sample size is small.

\begin{table}[htbp]
    \centering
    \caption{Simulated powers (\%) of the 
    score tests
    $S$, $S_e$, and $S_c$
     at a significance level of 5\%}
    \label{tab:sim-power}
\begin{threeparttable}
    \begin{tabular}{ccccccccccccccccc}\toprule
  &&\multicolumn{7}{c}{Scenario A}&\multicolumn{7}{c}{Scenario B}
	     \\
\cmidrule(lr){3-9}\cmidrule(lr){10-16}
$N_0$ &$w_0$&0.9&0.8&0.7&0.6&0.5&0.4&0.3
&0.9&0.8&0.7&0.6&0.5&0.4&0.3
	\\
	    \midrule
%
%
%
%

50&$S$
&15 &26&42&54&61&68&68
&9&12&16&20&25&30&37
	\\
&$S_e$
&12 &21&36&46&54&61&62
&5&7&9&11&14&16&17
	\\
&$S_c$
&10 &17&31&43&51&60&61
&5&6&9&11&14&17&20
	\\

100&$S$
&19 &37&60&77&85&88&88
&9&15&21&29&34&38&43
	\\
&$S_e$
&15&33&55&71&82&87&88
&5&10&15&21&26&30&34
	\\
&$S_c$
&13&31&52&69&80&87&88
&4&9&14&20&24&29&35
	\\

500&$S$
&51&93&99&100&100&100&100
&12&31&49&65&76&83&87
	\\
&$S_e$
&49&92&99&100&100&100&100
&10&29&46&63&76&84&87
	\\
&$S_c$
&48&91&100&100&100&100&100
&10&28&44&61&75&83&87
	\\
%
%
%
%
    \bottomrule
    \end{tabular}
\end{threeparttable}
\end{table}

\subsection{Comparison of abundance estimation}
\label{sec:est}

In each scenario,
we let $w_0$  be equal to 0.5, 0.7, and 0.9.
For abundance estimation,
we compare the proposed maximum EL estimator
$\widehat N_{\rm EL}$ ($\widehat N$ or $\widehat N_e$) with
the Horvitz--Thompson-type estimator $\widetilde N_{{\rm HT}}$
and the maximum EL estimator $\widetilde N$
under models without one-inflation.
In Table \ref{tab:sim-point}, we present
the simulated  means  and
relative mean square errors (the
ratios of the mean square errors to $N_0$)
of the three estimators
based on 2000 repetitions.

\begin{table}[h]
    \centering
    \caption{Simulated means
    and relative mean square errors (RMSEs)
    of the maximum EL estimators ($\widetilde N$ and
    $\widehat{N}_{EL}$) and the Horvitz--Thompson-type estimator
    $\widetilde{N}_{HT}$
    }
    \label{tab:sim-point}
\begin{threeparttable}
    \begin{tabular}{cccccccccccccc}\toprule

	&&&\multicolumn{3}{c}{$N_0 = 50$}&&
      \multicolumn{3}{c}{$N_0 = 100$}&&
        \multicolumn{3}{c}{$N_0 = 500$}
    \\

Scenario&	&$w_0$&0.5&0.7&0.9&&0.5&0.7&0.9&&0.5&0.7&0.9
\\
    \midrule

%
%

%
%

 A&	\multirow{1}{*}{Mean}&$\widetilde N$&
    113&75&56&&203&144&112&&963&699&552 \\

    &&$\widetilde N_{\rm HT}$&
    65&57&53&&111&108&104&&506&506&506
    \\

	&&$\widehat N_{\rm EL}$&
    56&52&50&&105&103&101&&503&503&502
    \\

    &\multirow{1}{*}{RMSE}&$\widetilde N$&
    139&20&3&&132&23&3&&441&82&7
    \\

    &&$\widetilde N_{\rm HT}$&
    42&6&2&&16&4&2&&1&2&3
    \\

	&&$\widehat N_{\rm EL}$&
    18&4&2&&10&3&2&&1&2&2
    \\

 B&	\multirow{1}{*}{Mean}&$\widetilde N$&
    115&83&63&& 196&143&115&&840&644&540\\

    &&$\widetilde N_{\rm HT}$&
    100&75&61&& 148&122&106&&547&526&505
    \\

	&&$\widehat N_{\rm EL}$&
    80&66&56&& 133&114&102&&530&515&498
    \\

    &\multirow{1}{*}{RMSE}&$\widetilde N$&
    204&80&23&& 181&48&11&&262&52&8\\

    &&$\widetilde N_{\rm HT}$&
    191&68&23&& 120&35&9&&35&16&7
    \\

	&&$\widehat N_{\rm EL}$&
    104&49&17&& 87&28&8&&28&14&7
    \\
    \bottomrule
    \end{tabular}
\end{threeparttable}
\end{table}

Clearly,
the maximum EL estimator $\widetilde N$ is severely upward-biased
when one-inflation is ignored in the regression models.
A possible explanation
for this observation is that the one-inflation
can be regarded as a limiting form of ``trap-shyness,'' and
ignoring it may make us falsely believe that the sample is skewed toward
those individuals with low capture probability.
This also explains why the probability of never being captured
is estimated to be higher than it
should be, and hence why the abundance estimates are
upward-biased.

After incorporating one-inflation parameters into the regression models,
it can be seen that
the proposed maximum EL estimator
$\widehat N_{\rm EL}$ always has smaller bias and RMSE
than the Horvitz--Thompson-type estimator $\widetilde N_{\rm HT}$,
especially when $N_0=50$ and 100.

Next, we compare the performance of
the EL ratio confidence interval $\mathcal{I}_{\rm EL}$
($\mathcal{I}_{1}$ or $\mathcal{I}_{1e}$)
and the Wald-type confidence interval
$\mathcal{I}_{\rm Wald}$ ($\mathcal{I}_2$ or $\mathcal{I}_{2e}$).
For this purpose, we present in Table \ref{tab:sim-interval}
the coverage probabilities of their two-  and
one-sided confidence intervals at a level of 95\%.
For the two-sided confidence intervals,
the coverage probability of $\mathcal{I}_{\rm EL}$
is very close to the nominal level, with the departure being at most 2\%.
Even though $\mathcal{I}_{\rm Wald}$ has an accurate
two-sided coverage probability when $N_0=500$,
it often produces severe undercoverage as $N_0$ is 100 or 50
with the largest undercoverage being 11\% in Scenario B.

\begin{table}[h]
    \centering
    \tabcolsep 4.6pt
    \caption{Simulated coverage probabilities (\%) of the
    EL ratio confidence interval $\mathcal{I}_{\rm EL}$
    and the Wald-type confidence interval $\mathcal{I}_{\rm Wald}$
    at a nominal level of 95\%}
    \label{tab:sim-interval}
\begin{threeparttable}
    \begin{tabular}{cccccccccccccccccc}\toprule

	&&&\multicolumn{3}{c}{$N_0 = 50$} &&
         \multicolumn{3}{c}{$N_0 = 100$}&&
        \multicolumn{3}{c}{$N_0 = 500$}
    \\

	Scenario&Type&$w_0$&0.5&0.7&0.9&&0.5&0.7&0.9&&0.5&0.7&0.9
\\
    \midrule

A&Two-sided&$\mathcal I_{\rm EL}$
&94&96&96&&94&95&96&&95&95&95 \\
&	&$\mathcal I_{\rm Wald}$
&89&91&90&&91&93&93&&94&95&96 \\
&Lower limit&$\mathcal I_{\rm EL}$
&96&96&98&&95&95&97&&95&95&95 \\
&	&$\mathcal I_{\rm Wald}$
&100&100&100&&100&100&99&&99&100&100 \\
&Upper limit&$\mathcal I_{\rm EL}$
&93&95&94&&94&95&95&&95&95&95 \\
&	&$\mathcal I_{\rm Wald}$
&87&88&87&&88&90&90&&91&92&93 \\

B&Two-sided&$\mathcal I_{\rm EL}$
&96&97&97&&96&97&97&&94&95&96\\
&	&$\mathcal I_{\rm Wald}$
&84&87&89&&90&91&91&&94&94&94\\
&Lower limit&$\mathcal I_{\rm EL}$
&96&97&98&&94&96&98&&92&94&96\\
&	&$\mathcal I_{\rm Wald}$
&100&100&100&&100&100&100&&100&100&100\\
&Upper limit&$\mathcal I_{\rm EL}$
&96&96&95&&96&97&95&&96&95&95\\
&	&$\mathcal I_{\rm Wald}$
&82&85&86&&88&88&89&&91&91&91\\

    \bottomrule
    \end{tabular}
\end{threeparttable}
\end{table}

\begin{figure}[htbp]
  \centering
  \includegraphics[width=5.5cm]{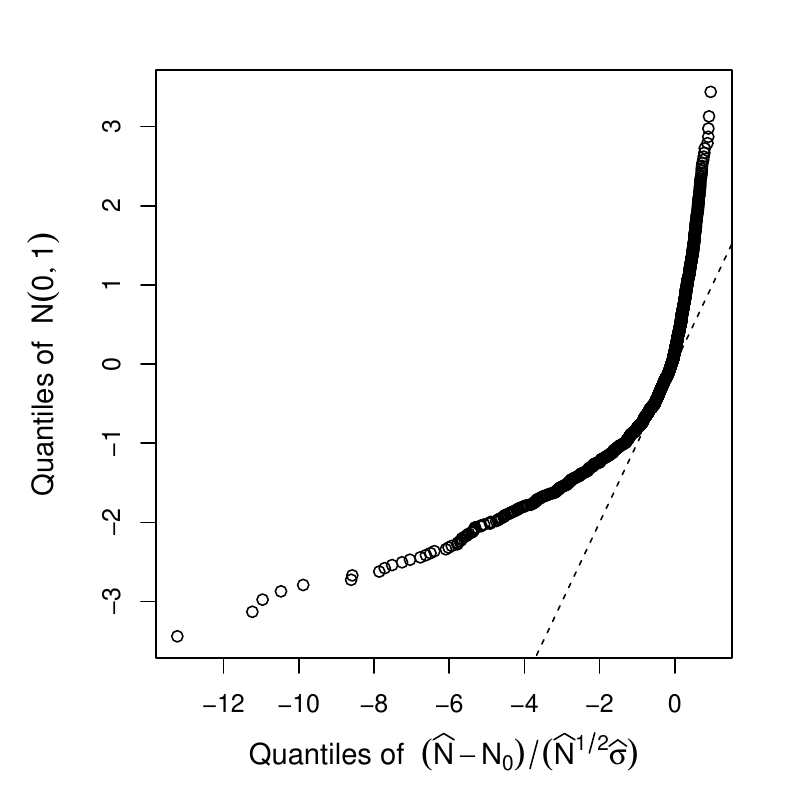}
  \includegraphics[width=5.5cm]{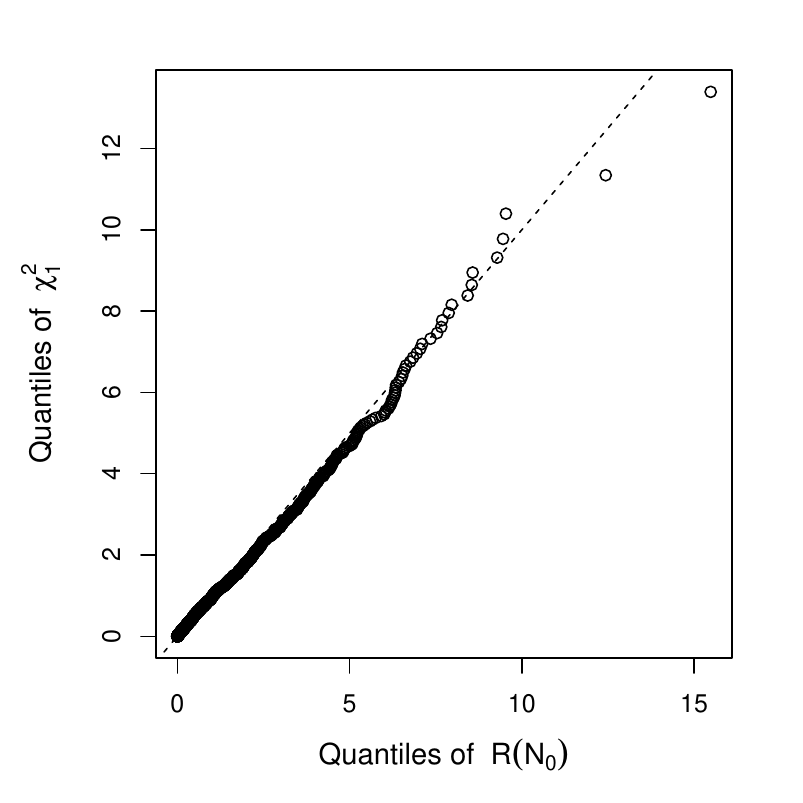}
  \caption{QQ plots of
  the pivotal statistic $(\widehat N - N_0)/(\widehat N^{1/2}\widehat\sigma)$ with
  the theoretical standard normal quantiles (left panel) and
  the empirical likelihood ratio $R(N_0)$ with
  the theoretical $\chi^2_1$ quantiles (right panel).
  In both panels, the dashed line is the identity line.
  }
  \label{fig:qqplot}
\end{figure}

Finally,
we display in Figure \ref{fig:qqplot}
the quantile--quantile (QQ) plots of
the pivotal statistic $(\widehat N - N_0)/\{ (\widehat N)^{1/2}\widehat\sigma\}$
and the EL ratio statistic $R(N_0)$
with their limiting distribution quantiles
in Scenario A with $N_0 = 50$ and $w_0=0.5$.
The plots for the other scenarios are similar
and are omitted.
Clearly, the quantiles of $R(N_0)$
are very close to those of the limiting standard chi-square distribution.
However, the sampling quantiles of  $(\widehat N - N_0)/\{ (\widehat N)^{1/2}\widehat\sigma\}$
are generally smaller than the quantiles of the limiting standard normal distribution.
This phenomenon would explain
why $\mathcal{I}_{\rm EL}$ has more accurate coverage probabilities
than $\mathcal{I}_{\rm Wald}$ and
why the lower limits of $\mathcal{I}_{\rm Wald}$
often produce overcoverage
but its upper limits often produce undercoverage.

\section{Real data analysis}\label{s:dat}
We further demonstrate the performance of
the proposed EL-based score tests and abundance estimation methods by
analyzing two real-life data sets:
prinia data
\citep{hwang2003estimation, liu2017maximum}
and drug user  data \citep{bohning2004estimating, bohning2009covariate}.
The prinia data were collected at the Mai Po Bird Sanctuary
in Hong Kong in 1993.
Across 17 weekly capture occasions,
164 yellow-bellied prinia were captured at least once,
and their wing lengths were measured.
The drug user data concerned
 female methamphetamine users in Bangkok in 2001.
  In total,
274 drug users had contacted  treatment institutions at least once,
and their age information was collected.
Here, the drug user was seen as being captured if she had contacted an institution once.
To  model  the processes of being captured,
we consider a binomial linear regression model and
a Poisson linear regression model as $f(y, x;\bbeta)$
for the prinia data and drug user data, respectively.

\subsection{Score tests for the existence of one-inflation}
We first test whether or not there exists one-inflation among the number of captures.
For this purpose, we perform
the EL-based score tests $S$
under the 
zero-truncated one-inflated model \eqref{eq:cap-prob-zero-trun}
and $S_e$ under the 
one-inflated zero-truncated model
\eqref{eq:cap-prob-inflate-positive-con}.
For the drug user data, both tests produce very small $p$-values,
$4\times 10^{-4}$ for $S$ and $7\times 10^{-5}$ for $S_e$,
which indicates that   one-inflation does exist
among the number of times that the drug users contacted the institutions.

The situation seems to be slightly different for the prinia data.
  The $p$-value of $S_e$ is 4.8\%,
which is very close to the significance level of 5\%.
However, the $p$-value of $S$ is 0.93\%,
which is strong evidence  for the existence of one-inflation
among the number of captures for the prinia data.
From this, we argue that
the EL-based score test $S$ under the model \eqref{eq:cap-prob-zero-trun},
is locally more powerful in detecting   one-inflation than the test $S_e$
under the model \eqref{eq:cap-prob-inflate-positive-con}.

\subsection{Abundance estimation}
In Table \ref{tab:data-est},
we present the estimation results for
the abundance of prinia and the total number of drug users.
Clearly, there are striking differences between
the maximum EL estimators $\widehat N_{\rm EL}$ and $\widetilde N$.
Compared with $\widehat N_{\rm EL}$,
ignoring the one-inflation problem usually
makes the maximum EL estimator $\widetilde N$ and
its standard error very large.
We apply the CL-based method of \cite{godwin2017estimation} to the
drug user data and obtainfor the Horvitz--Thompson-type estimator
$\widetilde N_{\rm HT}$ a value of 307 under the model
\eqref{eq:cap-prob-zero-trun}
and a value of 591 under the model \eqref{eq:cap-prob-inflate-positive-con}.
Consistent with the simulation results,
$\widetilde N_{\rm HT}$ is often larger than $\widehat N_{\rm EL}$.

With regard to the confidence intervals of the abundance,
we can see that the lower  and upper limits of
the Wald-type confidence interval $\mathcal{I}_{\rm Wald}$
are always lower than
their counterparts for the EL ratio confidence interval $\mathcal{I}_{\rm EL}$.
As discussed in the context of  our simulation studies,
the lower limit of the Wald-type confidence interval
often produces severe overcoverage and its upper limit often produces undercoverage.
In other words,
the lower limit may be much conservative and
the upper limit may lack  some confidence.
By contrast, the EL ratio confidence intervals could give more confidence.

\begin{table}
\centering
    \tabcolsep 7.2pt

\caption{Maximum EL estimators ($\widetilde N$,
 $\widehat N_{\rm EL}$, $\widehat w$, and $\widehat\omega_e$)
 and their standard errors, together with the EL ratio and Wald-type
 confidence intervals of the abundance,
 for the prinia data and drug user data}
\label{tab:data-est}
\begin{threeparttable}
\begin{tabular}{lcllllll}
\toprule
Data    &Model&$\widetilde N$& $\widehat N_{\rm EL}$  &$\mathcal{I}_{\rm EL}$
&$\mathcal{I}_{\rm Wald}$   &$\widehat w$ or $\widehat w_e$
\\
\midrule
Prinia&\eqref{eq:cap-prob-zero-trun}
&484 (94)\tnote{$\dag$}  &232 (49)  &[181,499]&[137, 327]& 0.66 (0.17)
\\
&\eqref{eq:cap-prob-inflate-positive-con}
 &484 (94)   &323 (75)  &[226,594]&[175, 470]& 0.58 (0.17)
\\

Drug user&\eqref{eq:cap-prob-zero-trun}
&2725 (518) &295 (40) &[276, 483] &[216, 373]&0.18 (0.18)
\\
&\eqref{eq:cap-prob-inflate-positive-con}
&2725 (518) &554 (192) &[340, 1442] &[177, 931]  &0.14 (0.07)
\\
\bottomrule
\end{tabular}
\begin{tablenotes}
\item[$\dag$] The estimates of the standard errors of the maximum EL estimators are shown in parentheses.
\end{tablenotes}
\end{threeparttable}
\end{table}

We may wonder which of the  models \eqref{eq:cap-prob-zero-trun}
and  \eqref{eq:cap-prob-inflate-positive-con} fits
 the prinia data and the drug user data
  better.
Table \ref{tab:data-est-freq} presents the observed and fitted marginal frequencies of
the numbers of captures in the real data sets.
Let us take the model \eqref{eq:cap-prob-zero-trun} and the prinia data as an example.
We regard the  observed and fitted marginal frequencies for  the prinia data
as a $2\times 5$ contingency table
and use Pearson's $\chi^2$ test statistic for independence  in the contingency table
to assess the goodness-of-fit of the  model \eqref{eq:cap-prob-zero-trun} for  the prinia data.
A smaller  Pearson's $\chi^2$ test statistic supports a better fit.
Table \ref{tab:data-est-freq} presents such test statistics for both of the models \eqref{eq:cap-prob-zero-trun}
and  \eqref{eq:cap-prob-inflate-positive-con}
and for both the  real data sets.
For  the prinia data,   Pearson's $\chi^2$ test statistic (2.33)
for the model \eqref{eq:cap-prob-zero-trun}  is smaller than
that (2.74) for the model \eqref{eq:cap-prob-inflate-positive-con}.
 This indicates that the zero-truncated one-inflated binomial regression model  \eqref{eq:cap-prob-zero-trun}
fits  the prinia data better.
Similarly.  the  one-inflated zero-truncated Poisson regression model   \eqref{eq:cap-prob-inflate-positive-con}
fits  the drug user data better.

\begin{table}
\centering
    \tabcolsep 14pt

\caption{Observed and fitted marginal frequencies, together with
the values of Pearson's $\chi^2$ test statistic, for
the number of captures in the prinia and drug user data}
\label{tab:data-est-freq}
\begin{threeparttable}
\begin{tabular}{cccccccc}
\toprule
Data    &Counts& 1  &2
&3  &4&5&
\\
\midrule
Prinia& Observed  &133 &24  &5&1& 1&
\\
&
 Fitted \eqref{eq:cap-prob-zero-trun}  &134 &23  &7&2& 0&$\chi^2=2.33$
\\
&
 Fitted \eqref{eq:cap-prob-inflate-positive-con} &137 &26  &7&2& 0&$\chi^2=2.74$
\\

Drug user & Observed &261 &10 &2 &1&
\\
&
Fitted \eqref{eq:cap-prob-zero-trun} &261 &9 &6 &5&&$\chi^2=5.58$
\\
&Fitted \eqref{eq:cap-prob-inflate-positive-con} &263 &10 &6 &4  & &$\chi^2=4.92$
\\
\bottomrule
\end{tabular}
\end{threeparttable}
\end{table}

\section{Discussion}\label{s:dis}

%


In this paper, we fit   capture--recapture data by
zero-truncated one-inflated and one-inflated zero-truncated
Poisson and binomial regression models.
More complicated count models have been investigated
in the literature to account for  unobserved heterogeneity
in the context of one-inflated capture--recapture data
\citep{godwin2017one, inan2018one, godwin2019one, bohning2021population}.
Taking  the unobserved heterogeneity into account,
we can alternatively define a geometric regression model
with the ``log'' link as
$$
f(y, \bx;\bbeta) = \left(\frac{ \mu(\bx; \bbeta) }
{1 + \mu(\bx; \bbeta)}\right)^y \frac{ 1 }{1 + \mu(\bx; \bbeta)},\quad
\mu(\bx; \bbeta) = 
\exp(\bbeta^\T x),
\quad \text{for } y=0, 1,2,\dots.
$$
Our semiparametric EL  estimation procedure
and the accompanying   EM algorithm can be straightforwardly
extended to deal with
the one-inflated geometric regression model.
In practice, some individual covariates are subject to missingness and
measurement errors. Ignoring these two factors may
lead to a biased abundance estimator \citep{stoklosa2019closed, liu2020maximum}.
Abundance estimation becomes much more challenging
when capture--recapture data are
contaminated missing data, one-inflation and zero-truncation.
We leave this work for future research.


\begin{acknowledgement}
The research  was supported by the China Postdoctoral Science Foundation (Grant 2020M681220),
the National Natural Science Foundation of China (11771144),
the State Key Program of National Natural Science Foundation of China (71931004 and 32030063),
the Development Fund for Shanghai Talents, and  the 111 project (B14019).
The first three authors contributed equally to this paper.
\end{acknowledgement}
\vspace*{1pc}

\noindent {\bf{Conflict of Interest}}

\noindent {\it{The authors have declared no conflict of interest.
}

%
%



\bibliography{mybibfile}



\vspace*{1pc}
\noindent {\bf{Supporting Information}}

\noindent {Additional supporting information may be found online in the Supporting Information section at the end of the article.}

\end{document}